# Charge transport in inhomogeneous quantum systems in weak electro-magnetic fields: two-time Green's function approach

Liudmila A. Pozhar[*]

Materials and Manufacturing Directorate (MLPS/MLBP), Air Force Research Laboratory, Wright-Patterson Air Force Base, OH 45433-7750

Western Kentucky University, Department of Chemistry

1 Big Red Way, Bowling Green, KY 42101

A first-principles theory of charge transport in spatially inhomogeneous quantum systems composed of any finite number of particles and subject to weak electro-magnetic fields is developed. Simple analytical expressions for the linear contributions to the quantum susceptibilities and conductivity tensors, and the charge and current densities of such systems are derived in terms of the two-time Green's functions. The obtained results can be applied to description of quantum charge transport in any finite system, including small semiconductor quantum dots, artificial molecules and atoms, etc.

# I. INTRODUCTION

Recent advances[1] in fundamental sciences and engineering made possible fabrication of nanoscale electronic devices[2] based on utilization of electronic properties of quantum dots (QDs) confined to spatial regions from tens to thousands of nanometers in linear dimensions. Smaller structures, such as nanoparticles, $C_{60}$, carbon nanotubes, nanowires, and other atomic and molecular clusters and assemblies, have also been fabricated and studied[3]. One of the major advantages of such artificial atomic scale systems is that their electronic properties can be suitably engineered to meet application requirements. The latest experimental efforts address fabrication of sub-nanostructures composed of atoms confined in well-characterized, atomic-dimension pores of solids (such as those of silicates fabricated using directed aqueous synthesis[4]) or lined to form two to three atomic layer regions in the space between low polymer $SiO_2$ or metal oxide blocks composed of several monomers and arranged in spatially symmetric structures by means of non-aqueous synthesis techniques[5]. Such small systems exemplify building units of future ultra compact, sub-nanoscale heterostructures, devices and chips. Understanding charge transport in such systems is extremely important and is being addressed experimentally[6] for the recent decade or so.

Following Alhassid[7], existing theoretical approaches to charge transport in finite systems can be roughly divided into several classes that include *quasi-* *classical*, *kinetic*, *Green's function*, *random-matrix (RMT)* and *supersymmetry methods*, according to the fundamental techniques from which they are derived. Within the framework of the quasi-classical approach one can derive, subject to some restrictive assumptions, an equation for the diffusive current of electrons in a perfect periodic potential in a crystal, where the motion of electrons is described by extended Bloch states. However, this description of conductivity breaks down already in the case of large mesosystems. System size effects manifest themselves in significant quantum contributions to electron motion, effects of charge-charge interactions, non-periodicity of the potential fields, preservation of the electron phase coherence after scattering from atomic or molecular boundaries, impurities, etc. One of the consequences of these size effects is that the charge transport in mesoscale and smaller systems is characterized in terms of the non-local, spatially dependent conductivity and susceptibility tensors, rather than via the constant conductivity and susceptibilities.

A more sophisticated phenomenological approach to charge transport in such systems was suggested originally by Landauer and Büttiker[8], and successfully applied[9] to relatively large nanosystems composed of thousands of atoms, such as large metallic QDs, quantum wells and wires (QWWs) that could be considered as weakly inhomogeneous systems or homogeneous systems weakly coupled to their environment. This approach has provided a valuable insight into the nature of charge transport in large nanosystems that has accounted for some of effects originating in finite nature of such systems. For example, such size effects shape the charge transport in metallic mesoscopic systems and in particular, manifest themselves via the Coulomb blockade that causes the conductance oscillations as a function of the gate voltage with a period that corresponds to an addition of a single electron to the system. In the case of metallic QDs weakly coupled to the leads via tunnel junctions, these size effects have been studied in the classical (the average energy level spacing $\Delta$ much smaller than temperature T) and quantum ($\Delta$>T) regimes by means of the Green's function technique[10] since the 1970s. This technique has also successfully described[11] the Coulomb staircase behavior of the conductance in the case of a metallic QD with a large number of excited states. At higher temperatures the coherence between the electrons in the leads and those in the dot may be ignored, and the description of charge transport reduces to that provided by the quantum kinetic approach[12]. While successful for large metallic QDs, Green's function-based approach has not been extended yet to include strongly inhomogeneous systems, such as semiconductor QDs, atomic and molecular clusters of few atoms, artificial atoms, etc. Attempts[13] to generalize this approach to account for interacting electrons using Keldysh's modification[14] of the Green's function technique was successful only in the zero-temperature limit, where electron-electron interactions can be viewed as elastic.

The RMT[15] (originally due to Wigner and Eisenbud) and its recent version formulated in terms of the effective, non-Hermitian Hamiltonians were applied[16] to ballistic QDs where the electron transport was defined by "chaotic" scattering from the dot boundaries weakly coupled to their environment. Using numerous assumptions, this approach recovers the Brett-Wigner resonance formula for the electron scattering amplitudes that are subsequently used in the Landauer formula to predict the zero-temperature conductance in the tunneling regime. The conductance at finite temperatures is calculated as a convolution of the zero-temperature conductance with the derivative of the Fermi-Dirac distribution (assumed constant). Many assumptions and fittings to known results make it difficult to evaluate the applicability of the RMT to small atomic clusters and similar systems where random scattering from the boundaries is unlikely to happen, boundary conditions are strongly non-uniform, and coupling to the confinements is likely to be strong. The RMT and supersymmetry[17] techniques have been used primarily for analysis of electronic transport statistics in large metallic QDs, due to the nature of the methods that deal with specific symmetries of the system Hamiltonians rather than with the Hamiltonians themselves. Several important features of QDs' conductance were discovered in the framework of these approaches, in particular the universality of the conductance fluctuations that do not depend on the average value of the conductance[18] and the dot geometry provided the electron scattering is random. However, quantum confinement effects reveal themselves in particular, via the influence of the leads[19] that are not adequately accounted for by the RMT and supersymmetry methods. Detailed analysis of these effects in the case of large metallic QDs has been provided by the Green's function formalism.

Considerable efforts to derive a self-consistent description of transport in small systems from the first non-equilibrium statistical mechanical principles[20] have led to development of a quantum kinetic equation-based approach[12, 21]. In the framework of this approach the complete many-body formulation of the problem reduces to the one-particle problem upon the use of an appropriate closure and boundary conditions to account for the quantum confinement effects. This approach has made possible a reliable description of charge transport in nanosystems, including relatively large semiconductor QDs and QWWs. Among other successes, the Landauer conductance formula was recovered in the framework of this approach as a contribution to the linear response conductance in the case of a confinement that entirely randomizes the electron phase. More importantly, this approach is related to important fundamental developments. Since 1960's it has been known[22] that the many-body problem can be mathematically formulated in terms of a system of coupled equations for the Bogolyubov-Tyablikov two-time Green's functions[23] (TTBTGF) (Dyson's equation is the simplest equation of this system[24]). A powerful method[25] for re-structuring and decoupling of this system has proved to be very successful for solving complicated problems of non-equilibrium statistical mechanics. It suggests a systematic and well-defined procedure (although there is no diagrammatic visualization of this procedure) that leads to calculation of the TTBTGFs in any desirable approximation. This procedure generalizes various projection operator techniques[26] developed and successfully applied to solve many demanding problems of non-equilibrium statistical mechanics and statistical physics. The TTBTGFs have a straightforward physical meaning of susceptibilities and can be used for direct calculations of various thermodynamics and transport properties.

In this paper the TTBTGF-based technique is used to develop a first-principles and tractable description of quantum charge transport in semiconductor QDs, small atomic clusters, artificial atoms and generally, in a spatially inhomogeneous system of any (finite or infinite) number of constitutive particles subject to a weak electro-magnetic field. For calculations of the linear (with respect to the fields) contributions to quantum susceptibilities and conductivity tensors at such conditions the theory of linear reaction is sufficient, while the full power of Zubarev-Tserkovnikov's method[22,26] has to be used for calculations of the corresponding non-linear contributions. These later calculations are postponed to future publications.

Recently, Lang *et al.*[27] developed a sophisticated technique designed specifically for calculations of the linear contribution to the quantum conductivity tensor of an inhomogeneous system in a weak electromagnetic field. While the suggested technique provides an insight into the problem of quantum charge transport in finite systems, it introduces unphysical and non-uniquely defined operators describing particle charge - field interactions, and does not lead to explicit analytical expressions for the quantum conductivity and susceptibility tensors. In the study presented below the TTBTGF-based method is used to solve the problem of linear quantum transport in a self-consistent manner, without a loss of generality or the use of *ad hoc* considerations. The resulting physically transparent and practical formulae represent the linear contributions to the susceptibilities and quantum conductivity tensors in terms of microscopic charge-charge and microcurrent - microcurrent equilibrium TTBTGFs of inhomogeneous systems that can be related to the equilibrium energy spectrum of charge carriers.

This work consists of eight sections. A brief introduction to the topic of this study is outlined in Sec. I. Sections II to V are devoted to the problem posing and derivation of the conservation equations for the quantum charge and current densities. Explicit expressions for the linear contributions to the generalized and dielectric susceptibility tensors, and the longitudinal sum rule are also derived in these sections. Sections VI and VII are focused on derivation of explicit expressions for the linear contributions to the longitudinal conductivity, magnetic susceptibility and the transversal conductivity. The developed approach and obtained results are discussed in Section VIII.

## II. PROBLEM FORMULATION

A system of $N$ particles each of which possesses the mass $m$ and charge $e$ is subject to a weak, time-dependent electromagnetic field of intensities $\mathbf{E}(\mathbf{r},t)$ and $\mathbf{H}(\mathbf{r},t)$:

$$\mathbf{E}(\mathbf{r},t) = -\frac{1}{c}\frac{\partial \mathbf{A}(\mathbf{r},t)}{\partial t} - \nabla_{\mathbf{r}}\varphi(\mathbf{r},t),$$
$$\mathbf{H}(\mathbf{r},t) = \mathbf{curl}\,\mathbf{A}(\mathbf{r},t), \qquad (1)$$

where bold symbols denote physical vectors; $\mathbf{r}$ and $t$ are a position vector and time, respectively; $\nabla_{\mathbf{r}}$ denotes the gradient operator with respect to $\mathbf{r}$; $c$ is the speed of light; and $\mathbf{A}(\mathbf{r},t)$ and $\varphi(\mathbf{r},t)$ are the vector and scalar potentials of the weak electromagnetic field in an arbitrary gauge, respectively. The external electromagnetic field is treated classically.

The total Hamiltonian $\mathcal{H}$ of the system has the form:

$$\mathcal{H} = \frac{1}{2m}\sum_{i=1}^{N}\left(\mathbf{p}_i - \frac{e}{c}\mathbf{A}(\mathbf{r}_i,t)\right)^2 + H_{\text{int}} + e\sum_{i=1}^{N}\varphi(\mathbf{r}_i,t), \qquad (2)$$

where $\mathbf{p}_i = \frac{\hbar}{i}\nabla_\mathbf{r}$ is the momentum operator of the particle $i$, and $H_{int}$ is the operator that includes interparticle interactions and interactions with the particles of the environment. The magnetization of the system is assumed to be small, so that the contribution to the Hamiltonian (2) due to interactions of particle spins with the magnetic field is neglected[28]. The second-quantized representation of the Hamiltonian (2) is

$$\mathcal{H} = \frac{1}{2m}\int d\mathbf{r}\,\psi^+(\mathbf{r})\left(\frac{\hbar}{i}\nabla_\mathbf{r} - \frac{e}{c}\mathbf{A}(\mathbf{r},t)\right)^2 \psi(\mathbf{r}) \\ + H_{int} + e\int d\mathbf{r}\,\psi^+(\mathbf{r})\psi(\mathbf{r})\varphi(\mathbf{r},t), \quad (3)$$

where $\psi^+(\mathbf{r})$ and $\psi(\mathbf{r})$ are the quantum field operators, related to the creation and annihilation operators of the charge carriers, $a^+_{\mathbf{k}\sigma}$ and $a_{\mathbf{k}\sigma}$, respectively, by the standard definitions:

$$\psi^+(\mathbf{r}) = \frac{1}{\sqrt{V}}\sum_{\mathbf{k},\sigma} e^{-i\mathbf{k}\mathbf{r}} \delta_{\sigma\sigma_z} a^+_{\mathbf{k}\sigma},$$
$$\psi(\mathbf{r}) = \frac{1}{\sqrt{V}}\sum_{\mathbf{k},\sigma} e^{i\mathbf{k}\mathbf{r}} \delta_{\sigma\sigma_z} a_{\mathbf{k}\sigma}, \quad (4)$$

with $\mathbf{k}$ being the wave vector, $\sigma_z$ denoting the value of the z-component of the spin, and $\delta_{\sigma\sigma_z}$ being the Kronecker delta.

The current density operator at a position **r** in the system at a time *t* is defined as the Fréchet derivative of the Hamiltonian with respect to the vector-potential $\mathbf{A}(\mathbf{r},t)$ of the external field, $\mathbf{j}(\mathbf{r},t) = -e \frac{\delta \mathcal{H}}{\delta \mathbf{A}(\mathbf{r},t)}$, and can be obtained from Eq. (3):

$$\mathbf{j}(\mathbf{r},t) = \frac{e\hbar}{2mi}\{\psi^+(\mathbf{r})\nabla_\mathbf{r}\psi(\mathbf{r}) - [\nabla_\mathbf{r}\psi^+(\mathbf{r})]\psi(\mathbf{r})\} - \frac{e^2}{mc}\mathbf{A}(\mathbf{r},t)\psi^+(\mathbf{r})\psi(\mathbf{r}). \tag{5}$$

Introducing the current density operator in the absence of the fields,

$$\mathbf{j}_0(\mathbf{r}) = \frac{e\hbar}{2mi}\{\psi^+(\mathbf{r})\nabla_\mathbf{r}\psi(\mathbf{r}) - [\nabla_\mathbf{r}\psi^+(\mathbf{r})]\psi(\mathbf{r})\}, \tag{6}$$

and the charge density operator

$$\rho(\mathbf{r}) = e\psi^+(\mathbf{r})\psi(\mathbf{r}) = en(\mathbf{r}), \tag{7}$$

where $n(\mathbf{r})$ denotes the number density of the charge carriers, one can rewrite the Hamiltonian (3) in the form:

$$\mathcal{H} = H_0 + H_1 + H_2, \tag{8}$$

where

$$H_0 = \frac{1}{2m}\int d\mathbf{r}\,\psi^+(\mathbf{r})\left[\frac{\hbar}{i}\nabla_\mathbf{r}\right]^2 \psi(\mathbf{r}) + H_{int} \tag{9}$$

is the unperturbed Hamiltonian in the absence of the electromagnetic field,

$$H_1 = -\frac{1}{c}\int d\mathbf{r}\, \mathbf{j}_0(\mathbf{r}) \cdot \mathbf{A}(\mathbf{r},t) + \int d\mathbf{r}\, \rho(\mathbf{r})\varphi(\mathbf{r},t), \tag{10}$$

is the perturbation Hamiltonian linear in the electromagnetic field potentials, and

$$H_2 = \frac{e}{2mc^2}\int d\mathbf{r}\rho(\mathbf{r})\mathbf{A}^2(\mathbf{r},t) \tag{11}$$

is the second-order perturbation Hamiltonian [the dots "·" in Eqs. (10) and (11) denote the inner product]. In view of the weakness of the magnetic field, the contribution (11) to the Hamiltonian (8) can be neglected, so that the resulting linearized Hamiltonian takes the form:

$$\mathcal{H} = H_0 + H_1 \tag{12}$$

For a system described by the Hamiltonian (12) one can obtain the expectation value of an observable using rigorous results of the quantum statistical theory of linear reaction of a system to weak perturbations (originally due to Kubo). [These results can be obtained via a trivial extension of this theory to include spatially inhomogeneous systems. Such an extension is possible due to a quasi-local nature[29] of the field operators (4).] Thus, the expectation value <**O**(**r**,t)> of an observable **O** can be expressed in terms of the equilibrium (or steady state) expectation value of **O**, $<\mathbf{O}(\mathbf{r})>_0$, [the corresponding state of the system is described by the Hamiltonian $H_0$ of Eq. (9)] and the equilibrium retarded, two-time temperature Bogoliubov-Tyablikov Green's function (ERTTBTGF) $<<\mathbf{O}(\mathbf{r},t)H_1(t')>>_0$ of the observable $\mathbf{O}$ and the perturbation Hamiltonian (10):

$$<\mathbf{O}(\mathbf{r},t)>=<\mathbf{O}(\mathbf{r})>_0 + \int_{-\infty}^{\infty} dt' <<\mathbf{O}(\mathbf{r},t)|H_1(t')>>_0. \tag{13}$$

In Eq. (13) $<\mathbf{O}(\mathbf{r},t)>=Tr(\wp\mathbf{O})$, where $Tr$ denotes the trace of an operator, and the statistical operator $\wp$ is the solution of the quantum Liouville equation with the Hamiltonian (12) and the equilibrium initial condition $\wp(t=-\infty) \equiv \wp_0 = e^{-\beta H_0}/Tre^{-\beta H_0}$, where $\beta=1/k_B T$ is the reciprocal temperature and $k_B$ is the Boltzmann constant. The equilibrium average value $<\mathbf{O}(\mathbf{r},t)>_0=Tr(\wp_0\mathbf{O})$ is defined with respect to the equilibrium statistical operator $\wp_0$, and the ERTTBTGF of any operators $\mathbf{O}$ and $\mathbf{K}$ is defined by the expression

$$<<\mathbf{O}(\mathbf{r},t)|\mathbf{K}(\mathbf{r}',t')>>_0 = \vartheta(t-t')\frac{1}{i\hbar}<[\mathbf{O}(\mathbf{r},t),\mathbf{K}(\mathbf{r}',t')]>_0, \tag{14}$$

where $\mathbf{r}'$ and $t'$ are the position vector and time, respectively, $\vartheta(t-t')$ is the step function and $[…,…]$ denotes the commutator.

Applying Eq. (13) to the non-equilibrium charge density operator $\rho(\mathbf{r},t) = e^{iH_0 t/\hbar} \rho(\mathbf{r}) e^{-iH_0 t/\hbar}$ and using Eq. (10), one can derive the charge conservation equation for a system in a state described by the Hamiltonian (12):

$$<\rho(\mathbf{r},t)> = \rho_0(\mathbf{r})$$
$$-\frac{1}{c}\int d\mathbf{r}' \int_{-\infty}^{\infty} dt' <<\rho(\mathbf{r},t)|\mathbf{j}_0(\mathbf{r}')>>_0 \cdot \mathbf{A}(\mathbf{r}',t') \quad (14)$$
$$+\int d\mathbf{r}' \int_{-\infty}^{\infty} dt' <<\rho(\mathbf{r},t)|\rho(\mathbf{r}')>>_0 \varphi(\mathbf{r}',t'),$$

where $\rho_0(\mathbf{r})$ is the charge density at the equilibrium or steady state. Similarly, applying Eq. (13) to the current density operator $\mathbf{j}(\mathbf{r},t)$ of Eq. (5), using Eq. (10) and noticing that in the absence of the field $<\mathbf{j}_0(\mathbf{r})>_0 = 0$, one can derive the current density conservation equation,

$$<\mathbf{j}(\mathbf{r},t)> = -\frac{1}{c}\int d\mathbf{r}' \int_{-\infty}^{\infty} dt' <<\mathbf{j}(\mathbf{r},t)|\mathbf{j}_0(\mathbf{r}')>>_0 \cdot \mathbf{A}(\mathbf{r}',t')$$
$$+\int d\mathbf{r}' \int_{-\infty}^{\infty} dt' <<\mathbf{j}(\mathbf{r},t)|\rho(\mathbf{r}')>>_0 \varphi(\mathbf{r}',t'), \quad (15)$$
$$-\frac{e}{mc}\rho_0(\mathbf{r})\mathbf{A}(\mathbf{r},t).$$

Note here, that the current density operator of Eq. (5) satisfies the microscopic charge conservation (or continuity) equation:

$$\frac{\partial \rho(\mathbf{r})}{\partial t} = -div\,\mathbf{j}(\mathbf{r},t).$$

# III. CONSERVATION EQUATIONS FOR THE SPACE-TIME FOURIER TRANSFORMS OF THE CHARGE AND CURRENT DENSITIES

The space-time Fourier transformations

$$<\mathbf{O}(\mathbf{r},t)> = \frac{1}{V}\sum_{\mathbf{k}} \int_{-\infty}^{\infty} d\omega \mathbf{O}(\mathbf{k},\omega) e^{i\mathbf{k}\cdot\mathbf{r}-i\omega t}$$

[where $V$ is the system volume, $\omega$ is the frequency and $\mathbf{k}$ is the wave vector] applied to Eqs. (14) and (15) lead to the corresponding conservation equations for the space-time Fourier transforms of the expectation values of the charge and current densities $\rho(\mathbf{k},\omega)$ and $\mathbf{j}(\mathbf{k},\omega)$, respectively:

$$\begin{aligned}\rho(\mathbf{k},\omega) = &\,en_0(\mathbf{k})\delta(\omega) \\ &-\frac{1}{c}\sum_{\mathbf{l}} <<\rho_{\mathbf{k}}|\,\mathbf{j}_{-\mathbf{l}}>>_{0,\omega} \cdot \mathbf{A}(\mathbf{l},\omega) \\ &+\sum_{\mathbf{l}} <<\rho_{\mathbf{k}}|\,\rho_{-\mathbf{l}}>>_{0,\omega} \varphi(\mathbf{l},\omega),\end{aligned} \qquad (16)$$

$$\mathbf{j}(\mathbf{k},\omega) = -\frac{e^2}{mc}\sum_{\mathbf{l}} n_0(\mathbf{k}-\mathbf{l})\mathbf{A}(\mathbf{l},\omega)$$
$$-\frac{1}{c}\sum_{\mathbf{l}} \ll \mathbf{j}_{\mathbf{k}} | \mathbf{j}_{-\mathbf{l}} \gg_{0,\omega} \cdot \mathbf{A}(\mathbf{l},\omega) \qquad (17)$$
$$+\sum_{\mathbf{l}} \ll \mathbf{j}_{\mathbf{k}} | \rho_{-\mathbf{l}} \gg_{0,\omega} \varphi(\mathbf{l},\omega),$$

where $\delta(\omega)$ denotes Dirac's delta function, the double brackets $\ll ... | ... \gg_{0,\omega}$ denote the space-time Fourier-transforms of the corresponding ERTTBTGFs, $n_0(\mathbf{k})$ is the space Fourier transform of the equilibrium charge number density [in the momentum representation $n_0(\mathbf{k}) = \frac{1}{V} < \sum_{\mathbf{q},\sigma} a^+_{(\mathbf{q}-\mathbf{k}),\sigma} a_{\mathbf{q},\sigma} >_0$], and the summations run over the wave vectors $\mathbf{l}$ ($-\infty \ll \mathbf{l} \ll \infty$). Thus, dependence of the Fourier transforms $\rho(\mathbf{k},\omega)$ and $\mathbf{j}(\mathbf{k},\omega)$ on the wave vectors is defined by the corresponding space-time Fourier transforms of the microscopic charge and current density operators. In the momentum representation these operators are:

$$\rho_{\mathbf{l}} = \frac{e}{\sqrt{V}} \sum_{\mathbf{q},\sigma} a^+_{\mathbf{q}-\mathbf{l},\sigma} a_{\mathbf{q},\sigma}, \qquad (18)$$

$$\mathbf{j}_{\mathbf{l}} = \frac{e\hbar}{m\sqrt{V}} \sum_{\mathbf{q},\sigma} (\mathbf{q}-\frac{\mathbf{l}}{2}) a^+_{\mathbf{q}-\mathbf{l},\sigma} a_{\mathbf{q},\sigma}, \qquad (19)$$

where the summation runs over the wave vectors **q** (-∞<<**q**<<∞) and $z$-components of the spin, $\sigma$.

The homogeneous system case is recovered from the conservation equations (16) and (17) upon a consideration that in this case each of the sums containing the Fourier-transforms of the ERTTTGFs reduces to only one non-zero term with **l=k**, and that the charge number density is constant. The equations (16) and (17) generalize the conservation equations of Refs. 22 to include the inhomogeneous system case.

### A. The generalized susceptibility and microcurrent - microcurrent ERTTBTGFs

Let's define the space-time Fourier transform $\chi_{\alpha\beta}(\mathbf{k},\mathbf{l},\omega)$ of the generalized susceptibility tensor $\chi$ by the following expression:

$$\chi_{\alpha\beta}(\mathbf{k},\mathbf{l},\omega) = \frac{e^2}{mc} n(\mathbf{k}-\mathbf{l})\delta_{\alpha\beta} + <<\mathbf{j}_{\mathbf{k}}^{\alpha}|\mathbf{j}_{-\mathbf{l}}^{\beta}>>_{0,\omega}, \qquad (20)$$

where the indices $\alpha$ and $\beta$ denote Cartesian components of the tensorial quantities, and $\delta_{\alpha\beta}$ is Kronecker's symbol. In the linear approximation studied here the Fourier transform of the microcurrent-microcurrent ERTTBTGF $<< j_{\mathbf{k}}^{\alpha} | j_{-\mathbf{l}}^{\beta} >>_{0,\omega}$ is a bilinear form of the vectors **k** and **l**. Therefore, it can be expressed in terms of the corresponding second rank tensors that are proportional to the unit matrix **I** and the tensor **kl** with coefficients that are scalar functions of the absolute values $|\mathbf{k}|\equiv k$, $|\mathbf{l}|\equiv l$, and the inner product $\mathbf{k}\cdot\mathbf{l}$, namely:

$$<< j_{\mathbf{k}}^{\alpha} | j_{-\mathbf{l}}^{\beta} >>_{0,\omega} = \chi^{lon}(\mathbf{k},\mathbf{l},\omega) \frac{k_{\alpha} l_{\beta}}{kl} + \left[ C(\mathbf{k},\mathbf{l},\omega) \delta_{\alpha\beta} - \frac{k_{\alpha} l_{\beta}}{kl} \right] \chi^{tr}(\mathbf{k},\mathbf{l},\omega), \quad (21)$$

where the space-time Fourier transform of the generalized scalar longitudinal susceptibility can be immediately identified,

$$\chi^{lon}(\mathbf{k},\mathbf{l},\omega) = \frac{<< \mathbf{k}\cdot \mathbf{j}_{\mathbf{k}} | \mathbf{l}\cdot \mathbf{j}_{-\mathbf{l}} >>_{0,\omega}}{kl}, \quad (22)$$

and the space-time Fourier transform of the generalized scalar transversal susceptibility $\chi^{tr}(\mathbf{k},\mathbf{l},\omega)$ and the scalar function $C(\mathbf{k},\mathbf{l},\omega)$ can be determined from the identity

$$\sum_{\alpha,\beta} \ll j_{\mathbf{k}}^{\alpha} | j_{-\mathbf{l}}^{\beta} \gg_{0,\omega} \left[ C(\mathbf{k},\mathbf{l},\omega)\delta_{\alpha\beta} - \frac{k_{\beta}l_{\alpha}}{kl} \right] =$$

$$\sum_{\alpha,\beta} \chi^{lon}(\mathbf{k},\mathbf{l},\omega) \frac{k_{\alpha}l_{\beta}}{kl} \left[ C(\mathbf{k},\mathbf{l},\omega)\delta_{\alpha\beta} - \frac{k_{\beta}l_{\alpha}}{kl} \right] \quad (23)$$

$$+ \left[ 3C^2(\mathbf{k},\mathbf{l},\omega) - 2\frac{\mathbf{k}\cdot\mathbf{l}}{kl}C(\mathbf{k},\mathbf{l},\omega) + \frac{(\mathbf{k}\cdot\mathbf{l})^2}{k^2 l^2} \right] \chi^{tr}(\mathbf{k},\mathbf{l},\omega),$$

that is obtained by using Eq. (21).

Substituting $\chi^{lon}(\mathbf{k},\mathbf{l},\omega)$ from Eq. (22) into Eq. (23), one can prove that the first term on the right hand side (r.h.s.) of Eq. (23) is equal to

$$C(\mathbf{k},\mathbf{l},\omega)\frac{\mathbf{k}\cdot\mathbf{l}}{k^2 l^2} \ll \mathbf{k}\cdot\mathbf{j}_{\mathbf{k}} |\mathbf{l}\cdot\mathbf{j}_{-\mathbf{l}} \gg_{0,\omega} - \frac{(\mathbf{k}\cdot\mathbf{l})^2}{k^3 l^3} \ll \mathbf{k}\cdot\mathbf{j}_{\mathbf{k}} |\mathbf{l}\cdot\mathbf{j}_{-\mathbf{l}} \gg_{0,\omega}.$$

By a convenient choice of the scalar function $C(\mathbf{k},\mathbf{l},\omega)$ (that does not effect a generality of this consideration),

$$C(\mathbf{k},\mathbf{l},\omega) = \frac{\mathbf{k}\cdot\mathbf{l}}{kl}, \quad (24)$$

this term can be set to zero. With this choice of $C(\mathbf{k},\mathbf{l},\omega)$, the r.h.s. of Eq. (23) becomes equal to $2\frac{(\mathbf{k}\cdot\mathbf{l})^2}{k^2 l^2}\chi^{tr}(\mathbf{k},\mathbf{l},\omega)$. The left hand side (l.h.s.) of Eq. (23) can be transformed further by the use of the identity $[\mathbf{k}\times\mathbf{j_k}]\cdot[\mathbf{l}\times\mathbf{j_{-l}}]=(\mathbf{k}\cdot\mathbf{l})(\mathbf{j_k}\cdot\mathbf{j_{-l}})-(\mathbf{k}\cdot\mathbf{j_{-l}})(\mathbf{l}\cdot\mathbf{j_k})$ (where the square brackets [...×...] denote the vector cross-product). In particular, dividing this identity by $kl$ and considering the ERTTBTGF $<<[\mathbf{k}\times\mathbf{j_k}]\cdot[\mathbf{l}\times\mathbf{j_{-l}}]>>_{0,\omega}$ (or formally, applying the operation $<<...>>_{0,\omega}$ to the above identity), one can prove that the l.h.s. of Eq. (23) is equal to $\frac{1}{kl}<<[\mathbf{k}\times\mathbf{j_k}]\cdot[\mathbf{l}\times\mathbf{j_{-l}}]>>_{0,\omega}$. Thus, Eq. (23) reduces to the following form:

$$\frac{1}{kl}<<[\mathbf{k}\times\mathbf{j_k}]\cdot[\mathbf{l}\times\mathbf{j_{-l}}]>>_{0,\omega}=2\frac{(\mathbf{k}\cdot\mathbf{l})^2}{k^2 l^2}\chi^{tr}(\mathbf{k},\mathbf{l},\omega). \qquad (25)$$

From Eq. (25) the scalar transversal susceptibility $\chi^{tr}(\mathbf{k},\mathbf{l},\omega)$ can be easily obtained:

$$\chi^{tr}(\mathbf{k},\mathbf{l},\omega)=\frac{kl}{2(\mathbf{k}\cdot\mathbf{l})^2}<<[\mathbf{k}\times\mathbf{j_k}]\cdot[\mathbf{l}\times\mathbf{j_{-l}}]>>_{0,\omega}. \qquad (26)$$

Substituting $C(\mathbf{k},\mathbf{l},\omega)$ from Eq. (24) into Eq. (21) one can determine the ERTTBTGF $<<j_\mathbf{k}^\alpha|j_\mathbf{-l}^\beta>>_{0,\omega}$:

$$<< j_{\mathbf{k}}^{\alpha} | j_{-\mathbf{l}}^{\beta} >>_{0,\omega} = \chi^{lon}(\mathbf{k},\mathbf{l},\omega)\frac{k_{\alpha}l_{\beta}}{kl} + \chi^{tr}(\mathbf{k},\mathbf{l},\omega)\left[\frac{(\mathbf{k}\cdot\mathbf{l})}{kl}\delta_{\alpha\beta} - \frac{k_{\alpha}l_{\beta}}{kl}\right], \quad (27)$$

where the scalar susceptibilities are defined by Eqs. (22) and (26). The expressions (22), (23) and (27) generalize the corresponding Zubarev's formulae of Refs. 22 to include the case of spatially inhomogeneous systems. They reduce to the corresponding Zubarev's expressions in the homogeneous system case.

### B. The longitudinal sum rule

Considerations similar to those discussed above in Sec. IIIA lead to explicit analytical expressions for the Fourier transforms of the ERTTBTGFs featuring in Eqs. (16) and (17). Thus, linearity of the considered approximation suggests that

$$<< j_{\mathbf{k}}^{\alpha} | \rho_{-\mathbf{l}} >>_{0,\omega} = << (\mathbf{k}\cdot\mathbf{j}_{\mathbf{k}}) | \rho_{-\mathbf{l}} >>_{0,\omega} \frac{k_{\alpha}}{k^2}, \quad (28)$$

$$<< \rho_{\mathbf{k}} | j_{-\mathbf{l}}^{\beta} >>_{0,\omega} = << \rho_{-\mathbf{k}} | (\mathbf{l}\cdot\mathbf{j}_{-\mathbf{l}}) >>_{0,\omega} \frac{l_{\beta}}{l^2}, \quad (29)$$

where the Fourier transforms $<< (\mathbf{k}\cdot\mathbf{j}_{\mathbf{k}}) | \rho_{-\mathbf{l}} >>_{0,\omega}$ and $<< \rho_{-\mathbf{k}} | (\mathbf{l}\cdot\mathbf{j}_{-\mathbf{l}}) >>_{0,\omega}$ are scalar functions of $k$, $l$ and $(\mathbf{k}\cdot\mathbf{l})$.

Using expressions Eqs. (28), (29) and (27), one can transform the conservation equations (16) and (17) to the form:

$$\rho(\mathbf{k},\omega) = e\sum_{\mathbf{l}} n_0(\mathbf{k})\delta(\omega)\delta_{\mathbf{kl}} - \frac{1}{c}\sum_{\mathbf{l}} \ll \rho_{\mathbf{k}} | (\mathbf{l}\cdot\mathbf{j}_{-\mathbf{l}}) \gg_{0,\omega} \frac{1}{l^2} \mathbf{l}\cdot\mathbf{A}(\mathbf{l},\omega)$$
$$+ \sum_{\mathbf{l}} \ll \rho_{\mathbf{k}} | \rho_{-\mathbf{l}} \gg_{0,\omega} \varphi(\mathbf{l},\omega),$$
(30)

$$\mathbf{j}(\mathbf{k},\omega) = -\frac{e^2}{mc}\sum_{\mathbf{l}} n_0(\mathbf{k}-\mathbf{l})\mathbf{A}(\mathbf{l},\omega)$$
$$-\frac{1}{c}\sum_{\mathbf{l}} \left\{ \chi^{lon}(\mathbf{k},\mathbf{l},\omega)\frac{\mathbf{kl}}{kl} + \chi^{tr}(\mathbf{k},\mathbf{l},\omega)\left[\frac{(\mathbf{k}\cdot\mathbf{l})}{kl}\mathbf{I} - \frac{\mathbf{kl}}{kl}\right] \right\}\cdot\mathbf{A}(\mathbf{l},\omega) \quad (31)$$
$$+ \sum_{\mathbf{l}} \ll (\mathbf{k}\cdot\mathbf{j}_{\mathbf{k}}) | \rho_{-\mathbf{l}} \gg_{0,\omega} \frac{\mathbf{k}}{k^2} \varphi(\mathbf{l},\omega),$$

where the generalized scalar susceptibilities are defined by Eqs. (22) and (26), and $n_0(\mathbf{k}-\mathbf{l})$ is the **k**-mode of the equilibrium charge number. Further calculations of the Fourier transforms of the ERTTBTGFs on the r.h.s. of Eqs. (30) and (31) can be achieved by the route similar to that suggested in Refs. 22. Namely, assuming pairwise additivity of interparticle interactions and using the momentum representation, one can prove that $\rho_{\mathbf{k}}$ commutes with the interaction Hamiltonian $H_{int}$ of Eq. (2), and that the commutator of $\rho_{\mathbf{k}}$ with the kinetic part of the Hamiltonian (2) is equal to $\hbar(\mathbf{k}\cdot\mathbf{j})$. Therefore, the equation of motion for the operator $\rho_{\mathbf{k}}$ takes the form:

$$\dot{\rho}_{\mathbf{k}} = \frac{1}{i\hbar}[\rho_{\mathbf{k}}, H_0] = -i(\mathbf{k}\cdot\mathbf{j}_{\mathbf{k}}), \quad (32)$$

where $H_0$ is the Hamiltonian (9), and $\dot{\rho}_{\mathbf{k}}$ denotes the Fourier-transform of the time derivative of the charge density operator. Using Eq. (32) one can calculate the commutator $\frac{1}{i\hbar}[\rho_{\mathbf{k}},\dot{\rho}_{-\mathbf{l}}]$ to obtain *the longitudinal sum rule* applicable to the case of spatially inhomogeneous systems:

$$\frac{1}{i\hbar}[\rho_{\mathbf{k}},\dot{\rho}_{-\mathbf{l}}] = \frac{e}{m}(\mathbf{l}\cdot\mathbf{k})\rho_{\mathbf{k}-\mathbf{l}}, \tag{33}$$

where in the momentum representation $\rho_{\mathbf{k}-\mathbf{l}} = \frac{e}{V}\sum_{\mathbf{q},\sigma} a^{+}_{\mathbf{q}-(\mathbf{k}-\mathbf{l}),\sigma} a_{\mathbf{q},\sigma}$. This sum rule reduces to the longitudinal sum rule of Refs. 22 specific to spatially homogeneous systems when $\mathbf{l}=\mathbf{k}$,

$$\frac{1}{i\hbar}[\rho_{\mathbf{k}},\dot{\rho}_{-\mathbf{k}}] = \frac{Ne}{mV}k^2, \tag{34}$$

with $N$ being the charge carrier number operator, $N = \sum_{\mathbf{q},\sigma} a^{+}_{\mathbf{q},\sigma} a_{\mathbf{q},\sigma}$, and the summations running over the wave vectors $\mathbf{q}$ and $z$-components of the spin, $\sigma$.

# IV. THE CHARGE CONSERVATION EQUATION IN TERMS OF THE ELECTRIC FIELD INTENSITY

Using Eqs. (32), (28) and (29), one can transform Eq. (30) to the form:

$$\rho(\mathbf{k},\omega) = e\sum_{\mathbf{l}} n_0(\mathbf{k})\delta(\omega)\delta_{\mathbf{kl}} + \frac{i}{c}\sum_{\mathbf{l}} \frac{1}{l^2} <<\rho_{\mathbf{k}}|\dot{\rho}_{-\mathbf{l}}>>_{0,\omega} (\mathbf{l}\cdot\mathbf{A}(\mathbf{l},\omega))$$
$$+ \sum_{\mathbf{l}} <<\rho_{\mathbf{k}}|\rho_{-\mathbf{l}}>>_{0,\omega} \varphi(\mathbf{l},\omega). \qquad (35)$$

Recovering the Fourier-image of $<<\rho_{\mathbf{k}}|\dot{\rho}_{-\mathbf{l}}>>_{0,\omega}$,

$<<\rho_{\mathbf{k}}|\dot{\rho}_{-\mathbf{l}}>>_{0,\omega} = \int_{-\infty}^{\infty} dt\, e^{i\omega t} <<\rho_{\mathbf{k}}|\dot{\rho}_{-\mathbf{l}}(t)>>_0$, integrating by parts on the

r.h.s. of this equation and noticing that $\frac{1}{i\hbar}<[\rho_{\mathbf{k}},\rho_{-\mathbf{l}}]>_0 = 0$, one can prove that

$$<<\rho_{\mathbf{k}}|\dot{\rho}_{-\mathbf{l}}>>_{0,\omega} = i\omega <<\rho_{\mathbf{k}}|\rho_{-\mathbf{l}}>>_{0,\omega}. \qquad (36)$$

Substituting this result into Eq. (35), one can recover the charge conservation equation in the form:

$$\rho(\mathbf{k},\omega) = e\sum_{\mathbf{l}} n_0(\mathbf{k})\delta(\omega)\delta_{\mathbf{kl}}$$
$$+ \sum_{\mathbf{l}} \frac{1}{l^2} <<\rho_{\mathbf{k}}|\rho_{-\mathbf{l}}>>_{0,\omega} \left\{ -\frac{\omega}{c}(\mathbf{l}\cdot\mathbf{A}(\mathbf{l},\omega)) + l^2 \varphi(\mathbf{l},\omega) \right\}. \qquad (37)$$

The term in the curly brackets in the r.h.s. of this equation can be easily found using the fact[30] that induced charges screen only the longitudinal component of the electrical field. In terms of the space-time Fourier transforms, this can be written as $\mathbf{l} \cdot \mathbf{D}(\mathbf{l},\omega) = \mathbf{l} \cdot \mathbf{E}(\mathbf{l},\omega)$, where $\mathbf{E}(\mathbf{l},\omega)$ is the space-time Fourier transform of the electric field $\mathbf{E}(\mathbf{r},t)$ from Eq. (1), and $\mathbf{D}(\mathbf{l},\omega)$ is the induced electric field. Therefore, applying the space-time Fourier transformations to Eq. (1) and finding the inner product $\mathbf{l} \cdot \mathbf{E}(\mathbf{l},\omega)$, one derives:

$$i\mathbf{l} \cdot \mathbf{D}(\mathbf{l},\omega) = -\frac{\omega}{c}(\mathbf{l} \cdot \mathbf{A}(\mathbf{l},\omega)) + l^2 \varphi(\mathbf{l},\omega). \tag{38}$$

The result (38) allows rewriting Eq. (37) in the form that does not contain explicitly the field potentials:

$$\rho(\mathbf{k},\omega) = e \sum_{\mathbf{l}} n_0(\mathbf{k})\delta(\omega)\delta_{\mathbf{kl}}$$
$$+ \sum_{\mathbf{l}} \frac{i}{l^2} \ll \rho_{\mathbf{k}} | \rho_{-\mathbf{l}} \gg_{0,\omega} \mathbf{l} \cdot \mathbf{D}(\mathbf{l},\omega). \tag{39}$$

This equation generalizes the corresponding Zubarev's result[22] to include the inhomogeneous system case, and reduces to Zubarev's charge conservation equation upon consideration that for a spatially homogeneous system $\ll \rho_{\mathbf{k}} | \rho_{-\mathbf{l}} \gg_{0,\omega} = \ll \rho_{\mathbf{k}} | \rho_{-\mathbf{k}} \gg_{0,\omega} \delta_{\mathbf{kl}}$.

## A. The polarization vector and the tensor of dielectric susceptibility

Following Nozieres and Pines, the induced field intensity $\mathbf{D}(\mathbf{r},t)$ linear in the external field $\mathbf{E}(\mathbf{k},\omega)$ can be written in a general form that satisfies the causality condition:

$$\mathbf{D}(\mathbf{r},t) = \int_0^\infty dt' \int d\mathbf{r}' [\delta(\mathbf{r}-\mathbf{r}')\delta(t-t')\mathbf{I} + \mathbf{F}(\mathbf{r}',t')] \cdot \mathbf{E}(\mathbf{r}-\mathbf{r}',t-t'), \tag{40}$$

where $\mathbf{F}(\mathbf{r}',t')$ is a (unknown) tensor of the second order, and the dummy variables of integration $\mathbf{r}'$ and $t'$ run over the space and time domains, respectively. Introducing formally the (unknown) second order Cartesian tensor of dielectric susceptibility $\boldsymbol{\varepsilon}(\mathbf{r}',t') = [\delta(\mathbf{r}-\mathbf{r}')\delta(t-t')\mathbf{I} + \mathbf{F}(\mathbf{r}',t')]$ and applying the space-time Fourier transformations to Eq. (40) one can derive a simple correlation between the Fourier transforms of the induced and applied electrical fields,

$$\mathbf{D}(\mathbf{l},\omega) = \boldsymbol{\varepsilon}(\mathbf{l},\omega) \cdot \mathbf{E}(\mathbf{l},\omega), \tag{41}$$

that holds for any physical system in a weak electromagnetic field. [In this equation $\boldsymbol{\varepsilon}(\mathbf{l},\omega)$ is the Fourier transform of the dielectric susceptibility tensor.] The difference between the fields $\mathbf{D}(\mathbf{r},t)$ and $\mathbf{E}(\mathbf{r},t)$ at a position $\mathbf{r}$ at a time $t$ is usually characterized by the polarization vector $\mathbf{P}(\mathbf{r},t)$,

$$\mathbf{P}(\mathbf{r},t) = \frac{1}{4\pi}[\ \mathbf{D}(\mathbf{r},t) - \mathbf{E}(\mathbf{r},t)\ ]. \tag{42}$$

The induced charge $\rho_{ind}(\mathbf{r},t) = -\nabla_{\mathbf{r}}\mathbf{P}(\mathbf{r},t)$ is defined by this polarization vector, so that the Fourier transform $\rho_{ind}(\mathbf{k},\omega)$ is

$$\rho_{ind}(\mathbf{k},\omega) = -\frac{i}{4\pi}\mathbf{k}\cdot[\ \mathbf{D}(\mathbf{k},\omega) - \mathbf{E}(\mathbf{k},\omega)\ ]. \tag{43}$$

This charge can be immediately found from Eq. (39):

$$\rho_{ind}(\mathbf{k},\omega) = \rho(\mathbf{k},\omega) - e\sum_{\mathbf{l}} n_0(\mathbf{k})\delta(\omega)\delta_{\mathbf{k}\mathbf{l}}$$
$$= \sum_{\mathbf{l}} \frac{i}{l^2} <<\rho_{\mathbf{k}}|\ \rho_{-\mathbf{l}}>>_{0,\omega} \mathbf{D}(\mathbf{l},\omega)\cdot\mathbf{l}. \tag{44}$$

Combining Eqs. (41), (43) and (44), and using considerations similar to those that have lead to explicit expressions for the generalized scalar susceptibilities (22) and (26), one can derive the dielectric susceptibility tensor in terms of the ERTTBTGFs $<<\rho_{\mathbf{k}}|\ \rho_{-\mathbf{l}}>>_{0,\omega}$. Thus, the linear in vectors $\mathbf{k}$ and $\mathbf{l}$ contribution to the tensor $\boldsymbol{\varepsilon}(\mathbf{l},\omega)$ that is of the major interest here can be written as follows:

$$\boldsymbol{\varepsilon}(\mathbf{k},\omega) = \sum_{\mathbf{l}}\left\{\varepsilon^{lon}(\mathbf{k},\mathbf{l},\omega)\frac{\mathbf{k}\mathbf{l}}{kl} + \varepsilon^{tr}(\mathbf{k},\mathbf{l},\omega)\left[\frac{(\mathbf{k}\cdot\mathbf{l})}{kl}\mathbf{I} - \frac{\mathbf{k}\mathbf{l}}{kl}\right]\right\},$$

where $\varepsilon^{lon}(\mathbf{k},\mathbf{l},\omega)$ and $\varepsilon^{tr}(\mathbf{k},\mathbf{l},\omega)$ are scalars that may depend only on the absolute values $k$, $l$ and the inner product $(\mathbf{k}\cdot\mathbf{l})$. In its turn, the field $\mathbf{E}(\mathbf{k},\omega)$ can be decomposed in two contributions that are parallel and orthogonal to the wave vector $\mathbf{k}$:

$$\mathbf{E}(\mathbf{k},\omega) = \frac{\mathbf{k}}{k^2}(\mathbf{k}\cdot\mathbf{E}(\mathbf{k},\omega)) + \frac{1}{k^2}[\mathbf{k}\times\mathbf{E}(\mathbf{k},\omega)\times\mathbf{k}] .$$

Substituting the above two expressions into Eq. (41) (where $\mathbf{l}$ is changed to $\mathbf{k}$) one can derive the following expression:

$$\begin{aligned}
\mathbf{k}\cdot\mathbf{D}(\mathbf{k},\omega) = \sum_{\mathbf{l}} & \left[\varepsilon^{lon}(\mathbf{k},\mathbf{l},\omega) - \varepsilon^{tr}(\mathbf{k},\mathbf{l},\omega)\right]\frac{k}{l}\times \\
& \left\{\frac{(\mathbf{l}\cdot\mathbf{k})}{k^2}(\mathbf{k}\cdot\mathbf{E}(\mathbf{k},\omega)) + \frac{\mathbf{l}\cdot[\mathbf{k}\times\mathbf{E}(\mathbf{k},\omega)\times\mathbf{k}]}{k^2}\right\} \\
& + \sum_{\mathbf{l}}\varepsilon^{tr}(\mathbf{k},\mathbf{l},\omega)\frac{(\mathbf{l}\cdot\mathbf{k})}{kl}(\mathbf{k}\cdot\mathbf{E}(\mathbf{k},\omega)).
\end{aligned} \qquad (45)$$

The induced magnetic field $\mathbf{B}(\mathbf{r},t)$ is related to the curl of the electric field, $\nabla_{\mathbf{r}}\times\mathbf{E}(\mathbf{r},t)$, by the Maxwell's equation $\nabla_{\mathbf{r}}\times\mathbf{E}(\mathbf{r},t) = -\frac{1}{c}\frac{\partial\mathbf{B}(\mathbf{r},t)}{\partial t}$. The corresponding equation for the space-time Fourier transforms reads:

$$[i\mathbf{k}\times\mathbf{E}(\mathbf{k},\omega)] = \frac{i\omega}{c}\mathbf{B}(\mathbf{k},\omega). \qquad (46)$$

Therefore, $\mathbf{l} \cdot [\mathbf{k} \times \mathbf{E}(\mathbf{k},\omega) \times \mathbf{k}] = \frac{\omega}{c} \mathbf{l} \cdot [\mathbf{B}(\mathbf{k},\omega) \times \mathbf{k}]$. Taking into account an already mentioned fact[30] that only the longitudinal component of the electrical fields defines the induced charge, one comes to a conclusion that the term proportional to $\frac{\omega}{c} \mathbf{l} \cdot [\mathbf{B}(\mathbf{k},\omega) \times \mathbf{k}]$ in the r.h.s. of Eq. (45) should be small and can be neglected in the linear approximation, thus leading to the following reduced form of Eq. (45):

$$\mathbf{k} \cdot \mathbf{D}(\mathbf{k},\omega) = \varepsilon(\mathbf{k},\omega)(\mathbf{k} \cdot \mathbf{E}(\mathbf{k},\omega)), \tag{47}$$

where the scalar dielectric susceptibility $\varepsilon(\mathbf{k},\omega)$ is defined by the "longitudinal" component of the tensorial dielectric susceptibility (that remains unknown), $\varepsilon(\mathbf{k},\omega) = \sum_{\mathbf{l}} \varepsilon^{lon}(\mathbf{k},\mathbf{l},\omega) \frac{(\mathbf{l} \cdot \mathbf{k})}{kl}$. Therefore, in the linear approximation with regard to the (weak) electromagnetic fields, the problem of determining the dielectric susceptibility reduces to determining a scalar quantity $\varepsilon(\mathbf{k},\omega)$ (albeit a function of all of the wave vectors).

Substituting the induced intensity of Eq. (47) into Eq. (43) one derives the following expression for the induced charge:

$$\rho_{ind}(\mathbf{k},\omega) = \frac{1}{4\pi} \left[ \frac{1}{\varepsilon(\mathbf{k},\omega)} - 1 \right] i\mathbf{k} \cdot \mathbf{D}(\mathbf{k},\omega). \tag{48}$$

Using this result in Eq. (44) one can obtain the desirable explicit expression for the space-time Fourier transform of the scalar dielectric susceptibility in terms of the space-time Fourier transform of the charge density - charge density ERTTBTGFs:

$$\varepsilon^{-1}(\mathbf{k},\omega) = 1 + 4\pi \sum_{\mathbf{l}} \frac{1}{l^2} <<\rho_{\mathbf{k}} | \rho_{-\mathbf{l}}>>_{0,\omega}. \qquad (49)$$

In the case of spatially homogeneous systems when $<<\rho_{\mathbf{k}} | \rho_{-\mathbf{l}}>>_{0,\omega} = <<\rho_{\mathbf{k}} | \rho_{-\mathbf{k}}>>_{0,\omega} \delta_{\mathbf{kl}}$, this expression reduces to its homogeneous system counterpart[22],

$$\varepsilon^{-1}(\mathbf{k},\omega) = 1 + \frac{4\pi}{l^2} <<\rho_{\mathbf{k}} | \rho_{-\mathbf{k}}>>_{0,\omega}. \qquad (50)$$

Note here, that Eq. (49) is obtained by neglecting the term proportional to $\frac{\omega}{c}\mathbf{l}\cdot[\mathbf{B}(\mathbf{k},\omega)\times\mathbf{k}]$ in Eq. (45), and thus is an approximate equation even in the framework of the linear response theory developed here, while its homogeneous system counterpart (50) is derived as an exact equation of the linear response theory of homogeneous systems. More general approximations leading to a derivation of explicit expressions for the tensorial dielectric susceptibility of Eq. (41) can be developed by solving the system of Eqs. (41), (43) and (44) without the neglect in Eq. (45) of the term proportional to $\mathbf{l}\cdot[\mathbf{B}(\mathbf{k},\omega)\times\mathbf{k}]$.

**C. The charge density conservation equation in terms of the field E(k,$\omega$)**

Using Eq. (49) one can transform Eq. (39) to the form that contains explicitly the electric field intensity $\mathbf{E}(\mathbf{k},\omega)$:

$$\rho(\mathbf{k},\omega) = e\sum_{\mathbf{l}} n_0(\mathbf{k})\delta(\omega)\delta_{\mathbf{kl}} + \sum_{\mathbf{l}} \frac{i}{l^2} \frac{<<\rho_{\mathbf{k}}|\rho_{-\mathbf{l}}>>_{0,\omega}}{\left\{1 + 4\pi\sum_{\mathbf{s}}\frac{1}{s^2}<<\rho_{\mathbf{l}}|\rho_{-\mathbf{s}}>>_{0,\omega}\right\}} \mathbf{l}\cdot\mathbf{E}(\mathbf{l},\omega). \quad (51)$$

This relation expresses the Fourier transform of the charge density in terms of the microscopic charge density - charge density ERTTBTGFs of an inhomogeneous system. Due to the use of an approximate Eq. (49) in derivation of Eq. (51), the latter equation is an approximation that holds only in weak electromagnetic fields. In the case of homogeneous systems when $<<\rho_{\mathbf{k}}|\rho_{-\mathbf{l}}>>_{0,\omega} = <<\rho_{\mathbf{k}}|\rho_{-\mathbf{k}}>>_{0,\omega}\delta_{\mathbf{kl}}$ Eq. (51) reduces to the charge conservation equation for homogeneous systems[22].

**V. THE CURRENT DENSITY CONSERVATION EQUATION**

The ERTTBTGFs of microcurrent appearing in Eqs. (22) and (31) can be obtained using Eq. (32):

$$\ll \mathbf{k} \cdot \mathbf{j_k} \mid \mathbf{l} \cdot \mathbf{j_{-l}} \gg_{0,\omega} = \ll \dot{\rho}_\mathbf{k} \mid \dot{\rho}_{-\mathbf{l}}) \gg_{0,\omega}, \tag{52}$$

$$\ll \mathbf{k} \cdot \mathbf{j_k} \mid \rho_{-\mathbf{l}} \gg_{0,\omega} = i \ll \dot{\rho}_\mathbf{k} \mid \rho_{-\mathbf{l}}) \gg_{0,\omega}. \tag{53}$$

Substituting the result (52) into the r.h.s. of Eq. (22), one can derive the longitudinal susceptibility in the form:

$$\chi^{lon}(\mathbf{k},\mathbf{l},\omega) = \frac{\ll \dot{\rho}_\mathbf{k} \mid \dot{\rho}_{-\mathbf{l}} \gg_{0,\omega}}{kl}. \tag{54}$$

Further transformations of the ERTTBTGFs in Eqs. (52) and (54) can be achieved by the use of the partial integration procedure similar to that that leads to Eq. (36). Thus, one can prove that:

$$\ll \dot{\rho}_\mathbf{k} \mid \dot{\rho}_{-\mathbf{l}} \gg_{0,\omega} = \omega^2 \ll \rho_\mathbf{k} \mid \rho_{-\mathbf{l}} \gg_{0,\omega} - \frac{1}{i\hbar}\langle[\rho_\mathbf{k},\dot{\rho}_{-\mathbf{l}}]\rangle_0, \tag{55}$$

and using the longitudinal sum rule (33), one can rewrite Eq. (55) as follows:

$$\ll \dot{\rho}_\mathbf{k} \mid \dot{\rho}_{-\mathbf{l}} \gg_{0,\omega} = \omega^2 \ll \rho_\mathbf{k} \mid \rho_{-\mathbf{l}} \gg_{0,\omega} - \frac{e^2}{m}(\mathbf{k}\cdot\mathbf{l})\, n_0(\mathbf{k}-\mathbf{l}), \tag{56}$$

where $n_0(\mathbf{k}-\mathbf{l})$ is the $(\mathbf{k}-\mathbf{l})$- mode of the equilibrium number density. Substituting the ERTTBTGF of Eq. (56) into Eq. (54) one can obtain the following expression for the space-time Fourier transform of the longitudinal susceptibility:

$$\chi^{lon}(\mathbf{k},\mathbf{l},\omega) = \frac{1}{kl}\left\{\omega^2 <<\rho_{\mathbf{k}} | \rho_{-\mathbf{l}})>>_{0,\omega} - \frac{e^2}{m} n_0 (\mathbf{k}-\mathbf{l})(\mathbf{k}\cdot\mathbf{l})\right\}. \quad (57)$$

The ERTTBTGF $<<\dot{\rho}_{\mathbf{k}} | \rho_{-\mathbf{l}}>>_{0,\omega}$ of Eq. (53) can be obtained by changing the sign of the r.h.s. of Eq. (36) and replacing the wave vectors $-\mathbf{l}$ by $\mathbf{k}$ and $\mathbf{k}$ by $-\mathbf{l}$, respectively:

$$<<\dot{\rho}_{\mathbf{k}} | \rho_{-\mathbf{l}}>>_{0,\omega} = -i\omega <<\rho_{\mathbf{k}} | \rho_{-\mathbf{l}}>>_0. \quad (58)$$

Substituting the Green's functions from Eqs. (53) and (58), and the longitudinal susceptibility (57) into Eq. (31) one obtains the following form of the current density conservation equation:

$$\begin{aligned}
\mathbf{j}(\mathbf{k},\omega) = & -\frac{e^2}{mc}\sum_{\mathbf{l}} n_0(\mathbf{k}-\mathbf{l})\mathbf{A}(\mathbf{l},\omega) \\
& -\frac{1}{c}\sum_{\mathbf{l}}\frac{1}{k^2 l^2}\left\{\omega^2 <<\rho_{\mathbf{k}} | \rho_{-\mathbf{l}})>>_{0,\omega} - \frac{e^2}{m} n_0 (\mathbf{k}-\mathbf{l})(\mathbf{k}\cdot\mathbf{l})\right\}(\mathbf{l}\cdot\mathbf{A}(\mathbf{l},\omega))\mathbf{k} \\
& +\sum_{\mathbf{l}}\frac{\omega}{k^2} <<\rho_{\mathbf{k}} | \rho_{-\mathbf{l}})>>_{0,\omega} \varphi(\mathbf{l},\omega)\mathbf{k} \\
& -\frac{1}{c}\sum_{\mathbf{l}}\chi^{tr}(\mathbf{k},\mathbf{l},\omega)\left\{\frac{(\mathbf{k}\cdot\mathbf{l})}{kl}\mathbf{A}(\mathbf{l},\omega) - \frac{1}{kl}(\mathbf{l}\cdot\mathbf{A}(\mathbf{l},\omega))\mathbf{k}\right\}.
\end{aligned} \quad (59)$$

Using the space-time Fourier transform of the continuity equation (A5) derived in Appendix, and Eqs. (37) and (59), one can prove the following equality:

$$\frac{1}{c}\chi^{tr}(\mathbf{k},\mathbf{l},\omega)\left\{\frac{(\mathbf{k}\cdot\mathbf{l})}{kl}+\frac{e^2}{m}n_0(\mathbf{k}-\mathbf{l})\right\}(\mathbf{k}\cdot\mathbf{A}(\mathbf{l},\omega)) =$$

$$\frac{1}{c}\left\{\chi^{tr}(\mathbf{k},\mathbf{l},\omega)+\frac{e^2}{m}\frac{(\mathbf{k}\cdot\mathbf{l})}{kl}n_0(\mathbf{k}-\mathbf{l})\right\}\frac{k^2}{kl}(\mathbf{l}\cdot\mathbf{A}(\mathbf{l},\omega)).$$

Using this equality and expressing the inner product $(\mathbf{l}\cdot\mathbf{A}(\mathbf{l},\omega))$ in terms of $(\mathbf{l}\cdot\mathbf{D}(\mathbf{l},\omega))$ from Eq. (38), one can transform the current density conservation equation (59) to the form:

$$\begin{aligned}\mathbf{j}(\mathbf{k},\omega) = & \sum_{\mathbf{l}}\frac{i\omega}{k^2 l^2}<<\rho_{\mathbf{k}}|\rho_{-\mathbf{l}})>>_{0,\omega}(\mathbf{l}\cdot\mathbf{D}(\mathbf{l},\omega))\,\mathbf{k} \\ & -\frac{1}{c}\sum_{\mathbf{l}}\left\{\left[\chi^{tr}(\mathbf{k},\mathbf{l},\omega)\frac{(\mathbf{k}\cdot\mathbf{l})}{kl}+\frac{e^2}{m}n_0(\mathbf{k}-\mathbf{l})\right]\right. \\ & \left.\times\left[\mathbf{A}(\mathbf{l},\omega)-\frac{1}{k^2}(\mathbf{k}\cdot\mathbf{A}(\mathbf{l},\omega))\,\mathbf{k}\right]\right\}.\end{aligned} \qquad (60)$$

The transversal component of the current density is defined by the induced magnetic field $\mathbf{B}(\mathbf{r},t)$ that in its turn, is defined by the transversal component of the vector-potential $\mathbf{A}(\mathbf{r},t)$ via the second equation of the system (1). The corresponding equation for the space-time Fourier-transforms reads:

$$\mathbf{B}(\mathbf{k},\omega) = [i\mathbf{k}\times\mathbf{A}(\mathbf{k},\omega)]. \qquad (61)$$

Note, the space-time Fourier-transform of the induced magnetic field $\mathbf{B}(\mathbf{r},t)$ is related to the corresponding space-time Fourier-transform of the curl of the electric field by the Maxwell's equation (46). The vector

$$\mathbf{B}_{\mathbf{k}}(\mathbf{l},\omega) = \frac{1}{k^2}[i\mathbf{k}\times i\mathbf{k}\times \mathbf{A}(\mathbf{l},\omega)] = \mathbf{A}(\mathbf{l},\omega) - \frac{(\mathbf{k}\cdot\mathbf{A}(\mathbf{l},\omega))}{k^2}\mathbf{k} \qquad (62)$$

featuring in the r.h.s. of Eq. (60) defines the component of the vector-potential which is orthogonal to the wave vector k. The vector on the r.h.s. of Eq. (62) can be transformed to the form:

$$\frac{1}{k^2}[i\mathbf{l}\times\mathbf{B}(\mathbf{l},\omega)] + \frac{(\mathbf{l}\cdot\mathbf{A}(\mathbf{l},\omega))}{l^2}\mathbf{l} - \frac{(\mathbf{k}\cdot\mathbf{A}(\mathbf{l},\omega))}{k^2}\mathbf{k},$$

where Eq. (61) has been used. The inner product of this vector with the wave vector **k** is zero, from which one can derive the following relation:

$$\mathbf{k}\cdot\mathbf{A}(\mathbf{l},\omega) = \frac{1}{l^2}\mathbf{k}\cdot[i\mathbf{l}\times\mathbf{B}(\mathbf{l},\omega)] + \frac{(\mathbf{l}\cdot\mathbf{k})}{l^2}(\mathbf{l}\cdot\mathbf{A}(\mathbf{l},\omega)). \qquad (63)$$

Using this equality and Maxwell's Eq. (46) in conjunction with Eq. (60), one can obtain a representation of the current conservation equation in the form convenient for further analysis:

$$\mathbf{j}(\mathbf{k},\omega) = i\omega \sum_{\mathbf{l}} \frac{<<\rho_{\mathbf{k}}|\rho_{-\mathbf{l}}>>_{0,\omega}}{k^2 l^2}(\mathbf{l}\cdot\mathbf{D}(\mathbf{l},\omega))\mathbf{k}$$

$$-\frac{1}{c}\sum_{\mathbf{l}}\frac{1}{l^2}\left[\chi^{tr}(\mathbf{k},\mathbf{l},\omega)\frac{(\mathbf{k}\cdot\mathbf{l})}{kl} + \frac{e^2}{m}n_0(\mathbf{k}-\mathbf{l})\right] \quad (64)$$

$$\times\left[\mathbf{I} - \frac{\mathbf{kk}}{k^2}\right]\cdot\{[i\mathbf{l}\times\mathbf{B}(\mathbf{l},\omega)] - \mathbf{l}(\mathbf{l}\cdot\mathbf{A}(\mathbf{l},\omega))\}.$$

This equation reduces to its homogeneous system counterpart[22] since in such a case the sums over **l** in the r.h.s. reduce to one term (with **l=k**) each.

Further progress toward expressing Eq. (64) in terms of the intensity $\mathbf{E}(\mathbf{k},\omega)$ is less straightforward. In the homogeneous system case the counterpart of Eq. (64) can be immediately transformed to an expression that does not include the vector-potential $\mathbf{A}(\mathbf{k},\omega)$ explicitly. In the inhomogeneous system case the presence of the summation over **l** in the second term in the r.h.s. of Eq. (64) does not allow similar exclusion of the vector-potential without an additional condition. A general form of such a condition widely used in electrodynamics is the Lorentz condition:

$$i\mathbf{l}\cdot\mathbf{A}(\mathbf{l},\omega) = \frac{i\omega}{c}\varphi(\mathbf{l},\omega). \quad (65)$$

This condition ascertains that the Maxwell's equations in vacuum are gauge-invariant. Using this condition in Eq. (38) one can express the longitudinal component of the vector-potential in terms of the longitudinal component of the intensity $\mathbf{D}(\mathbf{l},\omega)$:

$$i\mathbf{l}\cdot\mathbf{D}(\mathbf{l},\omega) = \frac{i\omega}{c}\left\{1 - \frac{c^2 l^2}{\omega^2}\right\} i\mathbf{l}\cdot\mathbf{A}(\mathbf{l},\omega).$$

Recovering $\mathbf{l}\cdot\mathbf{A}(\mathbf{l},\omega)$ from this equation and substituting it into Eq. (64) one obtains the current conservation equation in the form that does not contain explicitly the field potentials:

$$\begin{aligned}
\mathbf{j}(\mathbf{k},\omega) = &\sum_{\mathbf{l}} \frac{i\omega}{k^2 l^2} <<\rho_{\mathbf{k}}|\rho_{-\mathbf{l}}>>_{0,\omega} (\mathbf{l}\cdot\mathbf{D}(\mathbf{l},\omega))\mathbf{k} \\
&- \sum_{\mathbf{l}} \frac{1}{l^2}\left[\chi^{tr}(\mathbf{k},\mathbf{l},\omega)\frac{(\mathbf{k}\cdot\mathbf{l})}{kl} + \frac{e^2}{m}n_0(\mathbf{k}-\mathbf{l})\right] \\
&\times \left[\mathbf{I} - \frac{\mathbf{k}\mathbf{k}}{k^2}\right]\cdot\left\{\frac{1}{c}[i\mathbf{l}\times\mathbf{B}(\mathbf{l},\omega)] + \frac{i\mathbf{l}}{\omega\left(1 - \frac{c^2 l^2}{\omega^2}\right)}(\mathbf{l}\cdot\mathbf{D}(\mathbf{l},\omega))\right\}.
\end{aligned} \quad (66)$$

## VI. THE LONGITUDINAL CONDUCTIVITY

Using Maxwell's equation (46) and Eq. (47), (49) one can express the current density of Eq. (66) in the form that contains explicitly only the intensity $\mathbf{E}(\mathbf{l},\omega)$:

$$\mathbf{j}(\mathbf{k},\omega) = \frac{i\omega}{4\pi}[1-\varepsilon(\mathbf{k},\omega)]\frac{\mathbf{kk}}{k^2}\cdot\mathbf{E}(\mathbf{l},\omega)$$

$$-\sum_{\mathbf{l}}\frac{i}{\omega l^2}\left[\chi^{tr}(\mathbf{k},\mathbf{l},\omega)\frac{(\mathbf{k}\cdot\mathbf{l})}{kl} + \frac{e^2}{m}n_0(\mathbf{k}-\mathbf{l})\right] \quad (67)$$

$$\times\left[\mathbf{I}-\frac{\mathbf{kk}}{k^2}\right]\cdot\left\{[\mathbf{l}\times\mathbf{l}\times\mathbf{E}(\mathbf{l},\omega)] + \frac{\varepsilon(\mathbf{l},\omega)}{\left(1-\frac{c^2l^2}{\omega^2}\right)}\mathbf{l}(\mathbf{l}\cdot\mathbf{E}(\mathbf{l},\omega))\right\}.$$

This allows straightforward identification of the diagonal tensor of longitudinal conductivity in terms of the dielectric susceptibility,

$$\sigma^{lon}(\mathbf{k},\omega)\frac{\mathbf{kk}}{k^2} = \frac{i\omega}{4\pi}\{1-\varepsilon(\mathbf{k},\omega)\}\frac{\mathbf{kk}}{k^2}, \quad (68)$$

or explicitly, in terms of the charge-charge ERTTBTGFs using Eq. (49),

$$\sigma^{lon}(\mathbf{k},\omega) = \frac{i\omega\sum_{\mathbf{l}}\frac{1}{l^2}<<\rho_{\mathbf{k}}|\rho_{-\mathbf{l}}>>_{0,\omega}}{1+4\pi\sum_{\mathbf{l}}\frac{1}{l^2}<<\rho_{\mathbf{k}}|\rho_{-\mathbf{l}}>>_{0,\omega}}. \quad (69)$$

A possibility to establish closed explicit expressions [such as Eq. (69)] for the tensor of longitudinal conductivity depends entirely upon availability of a closed explicit expression for the dielectric susceptibility tensor. As discussed in Sec.

IVA, Eq. (49) used in derivation of Eq. (69) is an approximate equation that holds only in weak electromagnetic fields. An immediate consequence of this approximation is simplicity and diagonality of the tensor of longitudinal conductivity. Together with the linearization of the Hamiltonian (2) to the form (12), the above approximation of the dielectric susceptibility tensor is responsible for the fact that the linear contribution to the longitudinal conductivity (69) is defined completely by the microscopic charge density - charge density ERTTTGFs. Other approximations for the dielectric susceptibility tensor (still linear in the external fields) can be developed using Eqs. (41), (43) and (44). Such approximations will lead to more sophisticated (and more complex) explicit expressions for the longitudinal conductivity tensor that will also become non-diagonal. Development of such approximations is postponed to future publications.

Despite a simplicity of the approximate Eq. (49), the dielectric susceptibility tensor (49) and the tensor of the longitudinal conductivity of Eq. (69) still retain their dependence on the entire set of the microscopic charge density ERTTBTGFs. This dependence reflects spatial inhomogeneity of the system. In the homogeneous system case only one term (with **l=k)** in each of the sums over **l** in Eqs. (49) and (69) survives reducing these expressions to the form discussed in Refs. 22.

## VII. TRANSVERSAL CONDUCTIVITY

## A. The induced magnetic momentum and magnetic susceptibility

For the majority of experimentally studied cases, the generalized susceptibility $\chi^{tr}(\mathbf{k},\omega)$ of spatially homogeneous systems depends only on the wave vector $\mathbf{k}$ and is negative, so that the quantity $\{\chi^{tr}(\mathbf{k},\omega)+\frac{e^2}{m}n_0\}$ is small and describes only an insignificant diamagnetic effect derived originally by Landau. [This situation changes only in superconducting systems due to existence of the gap in the spectrum of the elementary excitations.] Similarly, one can expect that at standard conditions and in weak electromagnetic fields, for inhomogeneous systems with small magnetization the quantity $\{\chi^{tr}(\mathbf{k},\mathbf{l},\omega)\frac{(\mathbf{k}\cdot\mathbf{l})}{k\,l}+\frac{e^2}{m}n_0(\mathbf{k}-\mathbf{l})\}$ also remains small, albeit non-negligible. Thus, in the considered linear approximation with regard to the fields, the only manifestation of the spatial inhomogeneity effects is a dependence of this quantity on both $\mathbf{k}$ and $\mathbf{l}$ wave vectors.

Similar to electrical properties of a particle system in an external field that are described by the polarization vector, magnetic properties of a system in a magnetic field are characterized by the vector of induced magnetic moment $\mathbf{M}(\mathbf{r},t)$. This vector is proportional to the difference between the external and induced magnetic field intensities, so that its space-time Fourier-transform is:

$$\mathbf{M}(\mathbf{k},\omega) = \frac{1}{4\pi}[\mathbf{B}(\mathbf{k},\omega) - \mathbf{H}(\mathbf{k},\omega)]. \tag{70}$$

The magnetic moment defines the transversal contribution to the current density described by the second term in the r.h.s. of Eq. (67),

$$\mathbf{j}^{tr}(\mathbf{k},\omega) = c[i\mathbf{k} \times \mathbf{M}(\mathbf{k},\omega)]. \tag{71}$$

Upon a consideration similar to that leading to Eq. (41) for the induced electric field, one can show that in the linear approximation with regard to the field and due to the causality condition, the following correlation holds:

$$\mathbf{B}(\mathbf{k},\omega) = \boldsymbol{\mu}(\mathbf{k},\omega) \cdot \mathbf{H}(\mathbf{k},\omega), \tag{72}$$

where $\boldsymbol{\mu}(\mathbf{k},\omega)$ is the second-rank tensor of magnetic susceptibility. Using arguments similar to those discussed in Sec. IIIA in conjunction with the generalized susceptibility case, in the linear approximation this tensor can be represented by the following form:

$$\boldsymbol{\mu}(\mathbf{k},\omega) = \sum_{\mathbf{l}} \left\{ \mu^{lon}(\mathbf{k},\mathbf{l},\omega)\frac{\mathbf{kl}}{kl} + \mu^{tr}(\mathbf{k},\mathbf{l},\omega)\left[\frac{(\mathbf{k}\cdot\mathbf{l})}{kl}\mathbf{I} - \frac{\mathbf{kl}}{kl}\right]\right\}, \tag{73}$$

where $\mu^{lon}(\mathbf{k},\mathbf{l},\omega)$ and $\mu^{tr}(\mathbf{k},\mathbf{l},\omega)$ are scalar longitudinal and transversal magnetic susceptibilities that have to be determined and depend only on the absolute values $k$, $l$ and the inner product ($\mathbf{k}\cdot\mathbf{l}$). Substituting the result (73) into Eq. (72), one can rewrite Eq. (72) in the form:

$$\mathbf{B}(\mathbf{k},\omega) = \sum_{\mathbf{l}}\left\{\mu^{lon}(\mathbf{k},\mathbf{l},\omega) - \mu^{tr}(\mathbf{k},\mathbf{l},\omega)\right\}\frac{\mathbf{k}}{kl}(\mathbf{l}\cdot\mathbf{H}(\mathbf{k},\omega)) \\ + \sum_{\mathbf{l}}\mu^{tr}(\mathbf{k},\mathbf{l},\omega)\frac{\mathbf{k}\cdot\mathbf{l}}{kl}\mathbf{H}(\mathbf{k},\omega). \quad (74)$$

From this equation one can obtain the following correlation for the space-time Fourier-transforms of the curls of the intensities $\mathbf{H}(\mathbf{k},\omega)$ and $\mathbf{B}(\mathbf{k},\omega)$:

$$[i\mathbf{k}\times\mathbf{H}(\mathbf{k},\omega)] = \mu^{-1}(\mathbf{k},\omega)[i\mathbf{k}\times\mathbf{B}(\mathbf{k},\omega)], \quad (75)$$

where the scalar magnetic susceptibility $\mu(\mathbf{k},\omega)$ is

$$\mu(\mathbf{k},\omega) = \sum_{\mathbf{l}}\mu^{tr}(\mathbf{k},\mathbf{l},\omega)\frac{\mathbf{k}\cdot\mathbf{l}}{kl}. \quad (76)$$

For the following analysis it is convenient to use the current density conservation equation in the form (60). The second term in the r.h.s. of Eq. (60) (that is, the space-time Fourier transform of the transversal component of the induced current density) can be written in terms of the vector-potential $\mathbf{A}(\mathbf{k},\omega)$, and then expressed in terms of the magnetic susceptibility $\mu(\mathbf{k},\omega)$ as follows:

$$\mathbf{j}^{tr}(\mathbf{k},\omega) = \sum_{\mathbf{l}} \frac{i}{c k^2} \left[ \chi^{tr}(\mathbf{k},\mathbf{l},\omega) \frac{(\mathbf{k}\cdot\mathbf{l})}{kl} + \frac{e^2}{m} n_0(\mathbf{k}-\mathbf{l}) \right] [\mathbf{k}\times\mathbf{k}\times\mathbf{A}(\mathbf{l},\omega)]$$
$$= \frac{c}{4\pi}\left[1 - \mu^{-1}(\mathbf{k},\omega)\right][i\mathbf{k}\times\mathbf{B}(\mathbf{l},\omega)]. \tag{77}$$

The first line in the r.h.s. of Eq. (78) is obtained using Eqs. (71), (75) and (76), while the second line also uses Eq. (70). Comparing these two lines and using Eq. (61) again, one can establish the following explicit expression for the magnetic susceptibility $\mu(\mathbf{k},\omega)$:

$$\mu^{-1}(\mathbf{k},\omega) = 1 + \frac{4\pi}{c^2 k^2} \sum_{\mathbf{l}} \left[ \chi^{tr}(\mathbf{k},\mathbf{l},\omega) \frac{(\mathbf{k}\cdot\mathbf{l})}{kl} + \frac{e^2}{m} n_0(\mathbf{k}-\mathbf{l}) \right]. \tag{78}$$

### B. Explicit expression for the transversal conductivity

Using the second line in the r.h.s. of Eq. (77) for the transversal component of the space-time Fourier transform of the induced current density to replace the second term in the r.h.s. of Eq. (67), one can re-write Eq. (67) in the form:

$$\mathbf{j}(\mathbf{k},\omega) = \frac{i\omega}{4\pi}[1-\varepsilon(\mathbf{k},\omega)]\frac{\mathbf{k}\mathbf{k}}{k^2}\cdot\mathbf{E}(\mathbf{k},\omega)$$
$$+\frac{c}{4\pi}\left[1-\mu^{-1}(\mathbf{k},\omega)\right][i\mathbf{k}\times\mathbf{B}(\mathbf{k},\omega)],$$

or in view of Maxwell's equation (46),

$$\mathbf{j}(\mathbf{k},\omega) = \frac{i\omega}{4\pi}[1-\varepsilon(\mathbf{k},\omega)]\frac{\mathbf{k}}{k^2}\mathbf{k}\cdot\mathbf{E}(\mathbf{k},\omega)$$
$$-\frac{c^2 k^2}{4\pi i\omega}\left[1-\mu^{-1}(\mathbf{k},\omega)\right]\frac{1}{k^2}[\mathbf{k}\times\mathbf{k}\times\mathbf{E}(\mathbf{k},\omega)]. \tag{79}$$

From this equation the space-time Fourier transform of the scalar transversal conductivity $\sigma^{tr}(\mathbf{k},\omega)$ can be identified as the coefficient of $\frac{1}{k^2}[\mathbf{k}\times\mathbf{k}\times\mathbf{E}(\mathbf{k},\omega)]$, and thus related to the space-time Fourier transform of the scalar magnetic susceptibility,

$$\sigma^{tr}(\mathbf{k},\omega) = \frac{c^2 k^2}{4\pi i\omega}\{1-\mu^{-1}(\mathbf{k},\omega)\}. \tag{80}$$

As follows from Eq. (80), the transversal component of the conductivity tensor is completely defined by the equilibrium charge density and the microcurrent density – microcurrent density ERTTTGFs appearing in the explicit expression (78) for the magnetic susceptibility $\mu(\mathbf{k},\omega)$. With this result, one can reduce Eq. (67) to the *form* that formally coincides with that specific to the homogeneous system case,

$$\mathbf{j}(\mathbf{k},\omega) = \sigma^{lon}(\mathbf{k},\omega)\frac{\mathbf{kk}}{k^2}\cdot\mathbf{E}(\mathbf{k},\omega) \\ -\sigma^{tr}(\mathbf{k},\omega)\frac{1}{k^2}[\mathbf{k}\times\mathbf{k}\times\mathbf{E}(\mathbf{k},\omega)].$$  (81)

However, Eqs. (68) and (80) for the scalar longitudinal and transversal conductivities, respectively, included in the conservation equation (81) for the linear quantum current density in inhomogeneous systems differ significantly from their homogeneous case counterparts. In particular, the space-time Fourier-transforms of both quantum conductivities now depend upon the entire (infinite) set of the space-time Fourier transforms of the ETTBTGFs of the microscopic charge and current densities via Eqs. (69) and (78). Of course, simplicity of Eq. (81) originates from the use of the linear approximation that is justified by the weakness of the external electromagnetic field.

## C. Quantum conductivity of homogeneous systems

Noticing, that in the homogeneous system case the sums over **l** in the r.h.s. of Eq. (69) reduce to one term each (with **l=k**), one can immediately derive an expression for the linear contribution to the space-time Fourier transform of the longitudinal quantum conductivity of homogeneous systems in terms of the space-time Fourier transform of the microscopic charge density-charge density ERTTBTGF:

$$\sigma^{lon}(\mathbf{k},\omega) = \frac{\frac{i\omega}{k^2} \ll \rho_\mathbf{k} | \rho_{-\mathbf{k}} \gg_{0,\omega}}{1 + 4\pi \frac{1}{k^2} \ll \rho_\mathbf{k} | \rho_{-\mathbf{k}} \gg_{0,\omega}}. \quad (82)$$

An explicit expression for the linear contribution to the space-time Fourier transform of the transversal conductivity in terms of the space-time Fourier transform of the microcurrent - microcurrent ERTTBTGF can be obtained in a similar manner from Eqs. (26), (76), (78) and (80):

$$\sigma^{tr}(\mathbf{k},\omega) = \frac{1}{i\omega} \frac{\frac{1}{2k^2} \ll [\mathbf{k} \times \mathbf{j}_\mathbf{k}] \cdot [\mathbf{k} \times \mathbf{j}_{-\mathbf{k}}] \gg_{0,\omega} + \frac{e^2}{m} n_0}{1 + \frac{4\pi}{c^2 k^2} \left\{ \frac{1}{2k^2} \ll [\mathbf{k} \times \mathbf{j}_\mathbf{k}] \cdot [\mathbf{k} \times \mathbf{j}_{-\mathbf{k}}] \gg_{0,\omega} + \frac{e^2}{m} n_0 \right\}}. \quad (83)$$

Expressions (82) and (83) coincide with those derived in Refs. 22.

## VIII. DISCUSSION

The explicit expressions for the quantum susceptibilities and conductivities derived in this work are applicable to inhomogeneous systems of any nature and degree of inhomogeneity, including atomic and molecular clusters, quantum dots and wells, artificial atoms, etc. They reduce to their counterparts specific to the homogeneous system case when system inhomogeneity can be neglected. These expressions have been obtained within the framework of the linear response theory formulated in terms of ERTTBTGFs. The fact that derivation of such explicit expressions has become possible by the use of Zubarev-Tserkovnikov two-time Green's function formalism, while it has not been achieved via many other routes, including the recent technique[27] specifically developed to solve the problem of the quantum conductivity of strongly inhomogeneous systems, confirms yet again a strong potential and flexibility of Zubarev-Tsercovnikov's approach.

Reduction of the conservation equation (67) for the space-time Fourier transform of the linear quantum current density to Eq. (81) has become possible by approximating and keeping only the linear contribution with respect to the fields (or technically, the wave vectors) to the second term in the right hand side of Eq. (67). This linearization may not be applicable to strongly inhomogeneous systems subject to moderate and strong electromagnetic fields. Note also, that Eq. (31) for the space-time Fourier transform of the quantum current density does not use any assumptions or conditions concerning correlations between the scalar and vector field potentials and therefore, is the most general equation for the linear contribution to the space-time Fourier transform of the quantum current density in any inhomogeneous system with small non-linear effects. This equation provides a starting point for any possible developments, including the use of particular gauges for the field potentials. All results derived in this work reduce to their counterparts specific to the homogeneous system case if system inhomogeneity is negligibly small.

From the obtained results it follows that the linear contributions to the quantum charge transport in a system in a weak electromagnetic field can be entirely accounted for in terms of the microscopic charge density – charge density and microcurrent density – microcurrent density ERTTBTGFs. The ERTTBTGFs (and generally, TTBTGFs) of various observables have been intensively studied in numerous publications[22-25,31] available at present. It has been shown that the TTBTGFs of increasing complexity (*n*- operator TTBTGFs, where *n* runs from 2 to ∞) satisfy a system of coupled *algebraic* equations[25] that can be decoupled to any required accuracy by the use of the generalized continued fraction formalism. A general structure of the explicit expressions for the ERTTBTGFs so obtained is given in terms of the energy spectrum of the charge carriers specific to the equilibrium state of a system. This methodology has been intensively used to solve numerous general and particular problems of statistical mechanics and quantum field theory. The ERTTBTGFs appearing in the conservation equations for the charge and current densities, and in the explicit expressions for the quantum susceptibilities and conductivity derived above can be obtained by the use of this methodology (this work is postponed to future publications). Therefore, explicit calculations of the required ERTTBTGFs are entirely feasible, provided the charge carrier energy spectrum specific to the equilibrium state of the system is available.

The equilibrium energy spectra of the charge carriers can be derived theoretically solving the eigenvalue problem for the Schrödinger equation by existing analytical methods of equilibrium statistical mechanics and quantum field theory that are realized computationally. In the case of small atomic clusters of semiconductor atoms, artificial molecules, and other small atomic systems such computations can be routinely performed using existing Hartree-Fock/MCSCF- and DFT- based "quantum chemistry" software packages, such as GAMESS, GAUSSIAN-98, NWChem, etc. [Fast advances in computer hardware and software algorithm development will make such computations a matter of an engineering routine in the near future.]

Thus, the explicit formulae derived in this study reduce the problem of theoretical description of charge transport in inhomogeneous systems to the problem of theoretical predictions of the ERTTBTGFs that can be actually computed by existing computational means. [Note here, that practical applications of the formulae derived in this study may involve calculations of a large number of the corresponding ERTTBTGFs. This should not create a problem, provided the corresponding software is available.] The derived explicit expressions allow development of complementary software modules/packages that can be used in conjunction with existing quantum chemistry software to provide a fundamental tool for virtual (i.e., fundamental theory – based, computational) synthesis of sub-nanostructured materials and media with pre-designed transport properties for various applications in electronics and beyond. In particular, computational manipulations with the atomic cluster geometry, topology, composition and chemistry can be used to obtain data on possible electronic energy spectra of virtually pre-designed clusters[32]. These data can be further used to calculate the quantum susceptibilities, conductivities, and current and charge densities using the formulae reported above, to achieve the desirable electronic transport properties.

## ACKNOWLEDGMENTS

Support of the National Science Foundation (via the grant award DMR0340613), National Research Council, the Air Force Research Laboratory/Materials and Manufacturing Directorate, and the Air Force Office of Scientific Research of the U.S. Air Force is highly appreciated.

# APPENDIX

Using Eqs. (32) and (18) one can derive the following equation for the longitudinal component of the space-time Fourier transform of the current density:

$$\mathbf{k} \cdot \mathbf{j}(\mathbf{k},\omega) = -\frac{e^2}{mc} \sum_{\mathbf{l}} n_0(\mathbf{k}-\mathbf{l}) \mathbf{k} \cdot \mathbf{A}(\mathbf{l},\omega)$$
$$-\frac{i}{c} \sum_{\mathbf{l}} \ll \dot{\rho}_{\mathbf{k}} | \mathbf{j}_{-\mathbf{l}} \gg_{0,\omega} \cdot \mathbf{A}(\mathbf{l},\omega) \qquad (A1)$$
$$+ i \sum_{\mathbf{l}} \ll \dot{\rho}_{\mathbf{k}} | \rho_{-\mathbf{l}} \gg_{0,\omega} \varphi(\mathbf{l},\omega).$$

Recovering the Fourier-image of $\ll \dot{\rho}_{\mathbf{k}} | \mathbf{j}_{-\mathbf{l}} \gg_{0,\omega}$,

$\ll \dot{\rho}_{\mathbf{k}} | \mathbf{j}_{-\mathbf{l}} \gg_{0,\omega} = \int_{-\infty}^{\infty} dt\, e^{i\omega t} \ll \dot{\rho}_{\mathbf{k}}(t) | \mathbf{j}_{-\mathbf{l}} \gg_0$, and integrating by parts in the

r.h.s. of this equation, one can prove that

$$\ll \dot{\rho}_{\mathbf{k}} | \mathbf{j}_{-\mathbf{l}} \gg_{0,\omega} = -\frac{1}{i\hbar} < [\rho_{\mathbf{k}}, \mathbf{j}_{-\mathbf{l}}] >_0 - i\omega \ll \rho_{\mathbf{k}} | \mathbf{j}_{-\mathbf{l}} \gg_0. \qquad (A2)$$

Substituting the Fourier transform (A2) into Eq. (A1), using the conjugated Eq. (36) and Eq. (16), and noticing that in the absence of the field the system is uncharged, one can recover Eq. (A1) in the form:

$$\mathbf{k} \cdot \mathbf{j}(\mathbf{k},\omega) = -\frac{e^2}{mc} \sum_{\mathbf{l}} n_0(\mathbf{k}-\mathbf{l}) \mathbf{k} \cdot \mathbf{A}(\mathbf{l},\omega)$$
$$+ \frac{1}{c\hbar} \sum_{\mathbf{l}} <[\rho_{\mathbf{k}}, \mathbf{j}_{-\mathbf{l}}]>_0 \cdot \mathbf{A}(\mathbf{l},\omega) \qquad (A3)$$
$$+ \omega \rho(\mathbf{k},\omega).$$

Using Eqs. (18) and (19), one can express the commutator $[\rho_{\mathbf{k}}, \mathbf{j}_{-\mathbf{l}}]$ as follows:

$$\frac{1}{\hbar}[\rho_{\mathbf{k}}, \mathbf{j}_{-\mathbf{l}}] = \frac{e}{mV} \mathbf{k} \rho_{\mathbf{k}-\mathbf{l}}. \qquad (A4)$$

With this result and because of $<\rho_{\mathbf{k}-\mathbf{l}}>_0 = e n_0(\mathbf{k}-\mathbf{l})$, one can prove that the first two terms in Eq. (A3) sum up to zero, so that the space-time Fourier transform of the continuity equation for the charge density is recovered:

$$\mathbf{k} \cdot \mathbf{j}(\mathbf{k},\omega) = \omega \rho(\mathbf{k},\omega). \qquad (A5)$$

Therefore, in addition to the continuity equation for charge density operator, a similar continuity equation for its expectation value (i.e., the charge density) holds. It follows from Eq. (A5) that the induced charge density is determined by the induced current density. However, the converse statement is not correct, as only the longitudinal component of the induced current density is determined by the induced charge density.

---

* E-mail addresses: Liudmila.Pozhar@wpafb.af.mil; LPozhar@yahoo.com.

*New Methods in the Theory of Superconductivity* (Chapman & Hall, London, 1959); N. M. Plakida, in *Statistical Physics and Quantum Field Theory* (edited by N. N. Bogolyubov; Nauka, Moscow, 1973, pp. 205-240), etc.

[32] These and other aspects of such virtual synthesis of small atomic clusters with pre-designed electronic energy spectra are a subject of intensive research, and are further discussed in numerous publications, including L. A. Pozhar, *J. Chem. Phys.* (2004, in press); L. A. Pozhar, A. T. Yeates, F. Szmulowicz and W. C. Mitchel, *Mat. Res. Soc. Symp. Proc.* **788**, L11.40 (6p); L. A. Pozhar, *Mat. Res. Soc. Symp. Proc.* **789**, N3.10 (6p); **790**, P5.8 (6p). Theoretical calculations of pre-designed transport properties of sub-nanoscale systems can be found in L. A. Pozhar, V.F. de Almeida, and M.Z.-C. Hu, Ceramics Transactions, **137**, 101 (2003); J. M .D. MacElroy, L. A. Pozhar and S.- H. Suh, Colloids and Surfaces **A 187**-**A188**, 493 (2001); L. A. Pozhar, Phys. Rev. **E 61**, 1432 (2000); L. A. Pozhar and K. E. Gubbins, Int. J. Thermophys. **20**, 805 (1999); Phys. Rev. **E 56**, 5367 (1997), etc.